\documentclass[showpacs,amsmath,amssymb,twocolumn]{revtex4-2}
\usepackage{amssymb}
\usepackage{graphicx}

\usepackage{color}
\usepackage{ulem} 

\newcommand{\removed}[1]{}

\begin{document}

\title{Tunneling magnetoresistance in ensembles of ferromagnetic granules with exchange interaction and random easy axes of magnetic anisotropy.}

\author{Y.~M.~Beltukov}
\author{V.~I.~Kozub}
\author{A.~V.~Shumilin}
\affiliation{Ioffe Institute, 194021 St. Petersburg, Russia}
\author{N.~P.~Stepina}
\affiliation{Rzhanov Institute of Semiconductor Physics, 630090 Novosibirsk, Russia}

\begin{abstract}
We study the tunneling magnetoresistance in the ensembles of ferromagnetic granules with random easy axes of magnetic anisotropy taking into account the exchange interaction between granules. It is shown that due to the exchange interaction magnetoresistance is effectively decoupled from magnetization, i.e. the strongest negative magnetoresistance can be observed at the field where magnetization is almost saturated. Under some conditions, the sign of magnetoresistance can be reversed and tunneling magnetoresistance can become positive at certain magnetic fields.  Our theory agrees with measurements of magnetoresistance in ensembles of $\rm Fe$ granules in SiC${}_x$N${}_y$ matrix.
\end{abstract}

\maketitle

\section{Introduction}

The nanostructured ferromagnetic materials and, in particular, ensembles of ferromagnetic nanoparticles (granules) attract significant attention due to the diversity of their magnetic and conductive properties. It leads to numerous interesting physical phenomena including superparamagnetism and spin-glass-like behavior \cite{super} and possible applications in electronics, for example in magnetic memory and magneto-optics \cite{Nanomag,Anis}.

The tunneling magnetoresistance (TMR) is the essential property of ensembles of ferromagnetic granules and relates their conductive and magnetic properties \cite{InMae}. In hopping regime, the tunneling rates $w_{ij}$ depend to the mutual orientation of the magnetic moments of granules $i$ and $j$
\begin{equation} \label{MR0}
w_{ij} = w_{ij}^{(0)}(1 + P^2\cos\theta_{ij}).
\end{equation}
Here $\theta_{ij}$ is the angle between the magnetization of the granules,  $w_{ij}^{(0)}$ does not depend on orientations of magnetic moments. When the magnetic field leads to overall sample magnetization, the angles $\theta_{ij}$ decrease leading to negative magnetoresistance (MR).

Eq.~(\ref{MR0}) is significantly modified in the variable range hopping regime of conductivity (VRH). In this case the hopping occurs between the distant granules and includes co-tunneling through some number of Coulomb blockaded intermediate granules \cite{Zh-Shk}. Each co-tunneling process between intermediate granules $n$, $m$ adds a factor $(1 + P^2 \cos\theta_{nm})$ to the hopping probability \cite{Mae2, Koz-Sh}. Therefore, the hopping probability in VRH regime can be estimated as
\begin{equation} \label{MR1}
w_{ij} = w_{ij}^{(0)}(1 + P^2\cos\theta_{ij})^{N_{int}+1}
\end{equation}
where $N_{int}$ is the average number of intermediate granules in the hopping process. TMR in granules is discussed in the review \cite{GMR-rev}.

It is tempting to think that the magnetic field where TMR is observed is the field where the dependence of magnetization on magnetic field is the strongest. It can be shown that in an ensemble of independent identical granules the explicit relation exists  between MR and magnetization \cite{InMae, zhang1993, ferrari1997, Meil1}. In this case $\Delta R/ R \propto \langle \cos\theta_{ij} \rangle = \langle \cos\theta_{i} \rangle^2 \propto M_{tot}^2$.  Here $\theta_{i}$ is the angle between the magnetization of $i$-th granule and the external magnetic field. The averaged value of $\langle \cos\theta_{i} \rangle$ is proportional to the total sample magnetization $M_{tot}$. $\Delta R/ R$ is the relative correction to the resistance in the magnetic field. This relation between the MR and magnetization was reported in some experimental studies \cite{xiao1,xiao2,fan2014, balaev}. When the distribution of granule sizes is taken into account some deviations from $\Delta R/R \propto M_{tot}^2$ law are possible \cite{zhang1993, ferrari1997}. However, these deviations are not extremely strong and are most relevant at high magnetic field.

However, in a number of different granular ferromagnetic materials negative MR continue to grow even when magnetization is almost saturated. Similar behavior was observed in granular magnetic oxides \cite{Zeise2002,CrO22005,Fe3O42007,Fe3O42018}, nanoclusters of ferromagnetic metals \cite{CoAlN2007, C60Co2010,TiCrN} and in some other materials \cite{per2005,CaLaSr,FeMgO} including Mn-doped semiconductors \cite{GaAsrev}.
Usually, this MR is ascribed to existence of ``anti-boundaries'', the spins aligned against the main magnetization direction on the edge of granules \cite{Zeise2,Eerenstein}. Although in principle such spins can exist, their existence does not follow from any general law. It is doubtful that these reversed spins appear in all the materials where the discussed MR was measured.

Here we show that TMR itself can sometimes lead to magnetoresistance that is most strong in the fields when magnetization seems to saturate. It occurs due to the interplay of exchange interaction and magnetic anisotropy of the granules. When the exchange interaction is taken into account, the relation $\langle \cos\theta_{ij} \rangle = \langle \cos\theta_{i} \rangle^2$ is broken. It can lead to quite sophisticated dependencies of TMR on the applied magnetic field including the reversed sign of MR in certain magnetic fields. The discussed behavior of MR is observed when the anisotropy energy is large compared to the exchange energy. In this case the exchange energy significantly suppresses the TMR at the fields $H$ when the dependence $M_{tot}(H)$ is the strongest.
Note, that both the exchange interaction and anisotropy naturally appear in an array of ferromagnetic granules and exist in almost any material. It is important for our theory that the easy axes of anisotropy of different granules are random. It may occur due to non-spherical shape of granules \cite{shape1} or due to different crystallographic orientations in different granules.

We compare our theoretical results with  resent experiments obtained by us in the  SiC${}_x$N${}_y$:Fe granulated system. In this material we observed  the behavior of MR similar to that for several other mentioned granulated films: negative MR  linearly increases with magnetic field and does not saturate up to the high field while magnetization tends to saturate in small magnetic fields. The narrow peak of MR is observed with their width roughly match the range of strong magnetization change. The MR for samples with conductance being in hopping regime increase with decreasing the temperature demonstrating the increase of the hopping length at low temperatures.

The article is organized as follows. In Sec.~\ref{s:model} we discuss our model in details. In Sec.~\ref{s:num} we show the results of numeric simulation of our model. In Sec.~\ref{s:MF} we provide a simplified mean-field description of the model and compare it with numerical results. In Sec.~\ref{s:exp} we compare our theory with our recent experimental results on MR of Fe nanoclusters in SiCN matrix.

\section{Model of ferromagnetic granular material}
\label{s:model}

In this section we discuss the model that is applied to describe transport and magnetization of an array of ferromagnetic granules. The model includes all the effects crucial to our theory, i.e. the magnetic nature of the granules, anisotropy with random easy axis and exchange interaction. However, we tried to keep the model as simple as possible and did not include into the model several less significant features such as the distribution of granule sizes. 

\begin{figure}[htbp]
    \centering
        \includegraphics[scale=0.35]{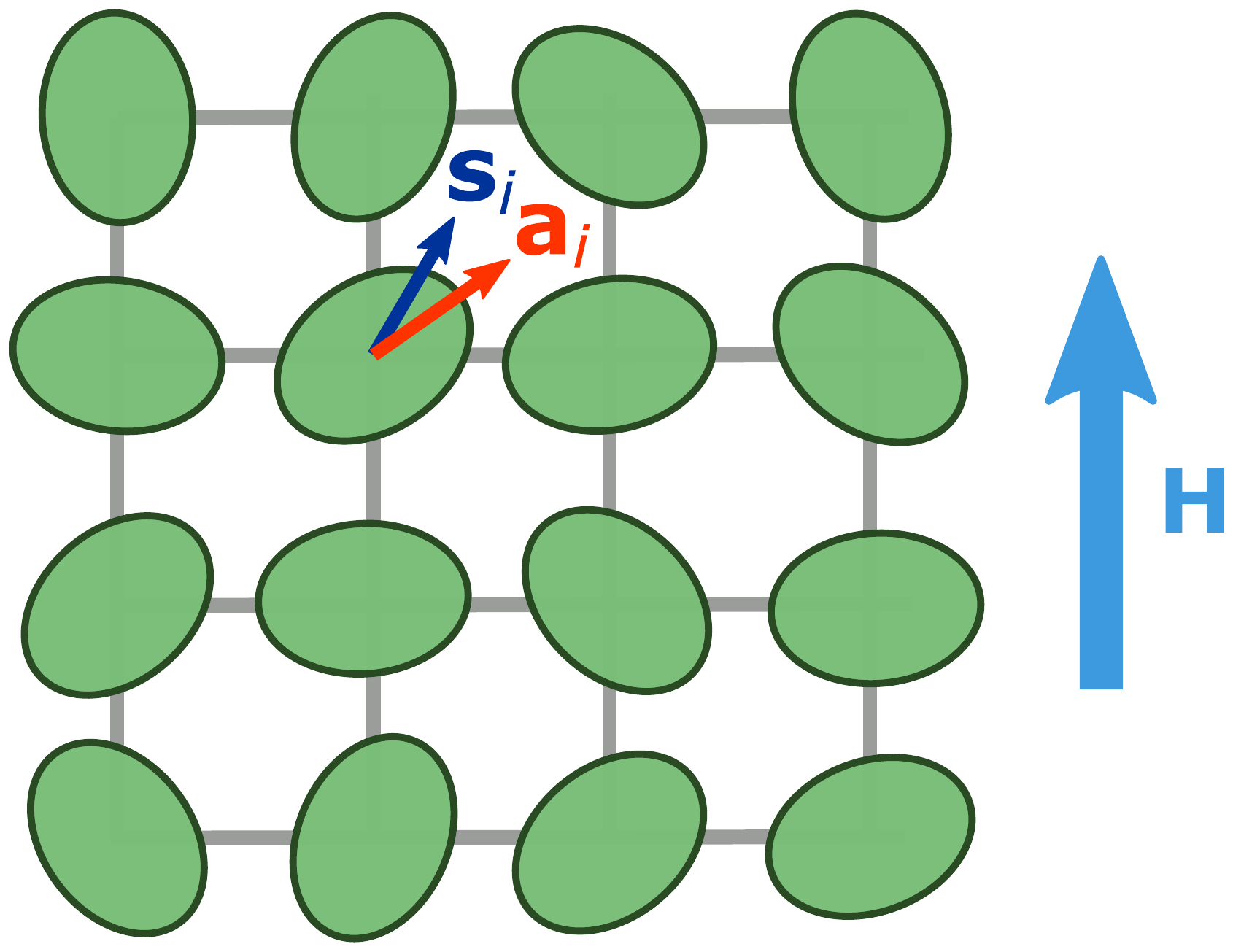}
        \caption{Model of the array of ferromagnetic granules. Each granule $i$ has easy axis ${\bf a}_i$. Its magnetization ${\bf s}_i$ can have other direction due to applied magnetic field and exchange interaction with neighbors.  }
    \label{fig:model}
\end{figure}

We describe the array of granules arranged on a square lattice (Fig.~\ref{fig:model}).
Each granule has random easy axis of magnetic anisotropy ${\bf a}_i$. The energy $K$  associated with the anisotropy is considered to be the same for all the granules. The magnetic moments ${\bf M}_i = M{\bf s}_i$ of all the granules have the same absolute value but different directions. Neighbor granules have ferromagnetic exchange interaction. The total energy of the system is as follows:
\begin{equation}\label{E-gen}
E = - K \sum_{i} ({\bf s}_i {\bf a}_i)^2  - M\sum_{i} {\bf s}_i {\bf H} - J \sum_{ij} {\bf s}_i{\bf s}_j.
\end{equation}
Here ${\bf s}_i$ is the unit vector in the direction of magnetization of the granule $i$.  $J$ is the exchange energy.
This model of granular ferromagnetic system was discussed in \cite{MaoMonte, Ilu-Koz, Bel-Koz}, with dipole-dipole interaction sometimes added to the energy $E$.
In \cite{Monte1998, Monte2004, Landau2010,  Monte2019} a similar model without ferromagnetic exchange coupling but with dipole-dipole interaction was treated numerically.

The magnetization of the system is equal to ${\bf M}_{tot} = M\sum_i {\bf s}_i$. In a macroscopic system it is always directed along the magnetic field and is equal to $N_{gr}M\langle\cos\theta_i \rangle$ where $N_{gr}$ is the number of granules and $\theta_i$ is the angle between magnetic field and ${\bf s}_i$.
Transport in a hopping system can be described by Miller-Abrahams resistor network \cite{MA, Efr-Sh}.  The conductivities of its resistors are proportional to the hopping rates $w_{ij}$. According to Eq.~(\ref{MR1}), the effect of the applied magnetic field on
these conductivities is controlled by $\cos\theta_{ij}$. Usually in a hopping system the distribution of the conductivities is exponentially broad and only rare critical resistors of the percolation theory are relevant to the macroscopic resistivity of the sample. Therefore, to calculate the MR one should average $\cos\theta_{ij}$ over all the critical resistors $ij$.  However, we presume that the distribution of $\cos\theta_{ij}$ is almost the same for critical and non-critical resistors. Therefore, we average $\cos\theta_{ij}$ over all the neighbor pairs of granules to calculate TMR.
\begin{equation} \label{MR-gen}
\frac{\Delta R}{R(0)} =  - {\cal P} \big(\langle\cos\theta_{ij} \rangle (H) - \langle\cos\theta_{ij} \rangle (0) \big).
\end{equation}
Here $\Delta R = R(H) - R(0)$ is the correction to the system resistivity due to the magnetic field. It is considered to be small compared to $R(0)$. ${\cal P} = P^2 (N_{int} +1)$ is the constant that controls the magnitude of the effect of granule magnetizations on conductivities. It is related to the polarization of material $P$ and the average number $N_{int}$ of intermediate granules in the hop. In VRH regime $N_{int}$ grows with decreasing temperature.

\section{Numeric simulation}
\label{s:num}

There are different methods to simulate a thermal equilibrium of magnetic granules. The direct method is to simulate the granular magnetic system using the Stochastic Landau-Lifshitz equation \cite{Romeo-2008, Leliaert-2017}. However, it requires a simulation with a small timestep. Another issue of this method is exponentially slow relaxation between two minima separated by the barrier caused by anisotropy.

In this paper we study a thermal equilibrium of a system rather than the magnetization dynamics. In this case we can use the Metropolis algorithm~\cite{Metropolis} to describe a system at a given temperature $T$. We simulate a periodic sample with $N=30\times30$ granules. For each step we make a random perturbation $\boldsymbol{\xi}_i^{(n)}$ of each magnetization direction using the following Markov scheme:
\begin{equation}\label{Mark1}
    {\bf s}_i^{(n + 1)} = \frac{{\bf s}_i^{(n)} + \boldsymbol{\xi}_i^{(n)}}{|{\bf s}_i^{(n)} + \boldsymbol{\xi}_i^{(n)}|}.
\end{equation}
This equation ensures that ${\bf s}_i^{(n + 1)}$ is unit vector.
Following the Metropolis algorithm, the new state ${\bf s}_i^{(n + 1)}$ is accepted with probability
\begin{equation} \label{MC_p}
    p = \min\left[\exp\left(-\frac{E^{(n+1)} - E^{(n)}}{T}\right), 1\right].
\end{equation}
Otherwise, the state remains unchanged: ${\bf s}_i^{(n + 1)} = {\bf s}_i^{(n)}$.
In Eq.~(\ref{MC_p}) $E^{(n)}$ is the energy of the system on step $n$ and $E^{(n+1)}$ is the energy of the perturbed state.

The random perturbation $\boldsymbol{\xi}_i^{(n)}$ was chosen as an uncorrelated random Gaussian vector
\begin{equation}
    \langle\xi_{i\alpha}^{(n)} \xi_{j\beta}^{(m)}\rangle = \sigma^2\delta_{nm}\delta_{ij} \delta_{\alpha\beta}.
\end{equation}
The variance $\sigma^2$ was chosen to meet the balance between a large step and a reasonable acceptance rate: $\sigma^2=0.16 T/N$. For a small temperature $T$, there is a barrier for magnetization of each granule due to anisotropy $K$. To increase the probability to hop over this barrier we make the following trick. With  a small probability $p_u$ for each granule $i$ at each step $n$ we take the magnetization direction ${\bf s}_i^{(n + 1)}$ from uniform distribution on unit sphere instead of the perturbation described by Eq.~(\ref{Mark1}). In this case the probability to get over the barrier is proportional to $p_u T$ instead of a small activation exponent $\exp(-T/K)$. Numerical simulation shows that $p_u = 1/N$ is close to an optimal value.

To obtain a system at different temperatures we start from some temperature $T=T_0\sim K$, which is comparable with a typical barrier energy of the system. Then we gradually decrease the temperature of the system using the robust logarithmic annealing protocol: $T^{(n)} = T_0/\ln(e+b n)$ where $n$ is the step number \cite{Geman-1984}. For system under consideration, the parameter $b=10^{-5}$ was small enough to obtain a thermal equilibrium for $T \geq 0.01 K$.

\begin{figure*}[htbp]
    \centering
        \includegraphics[scale=0.8]{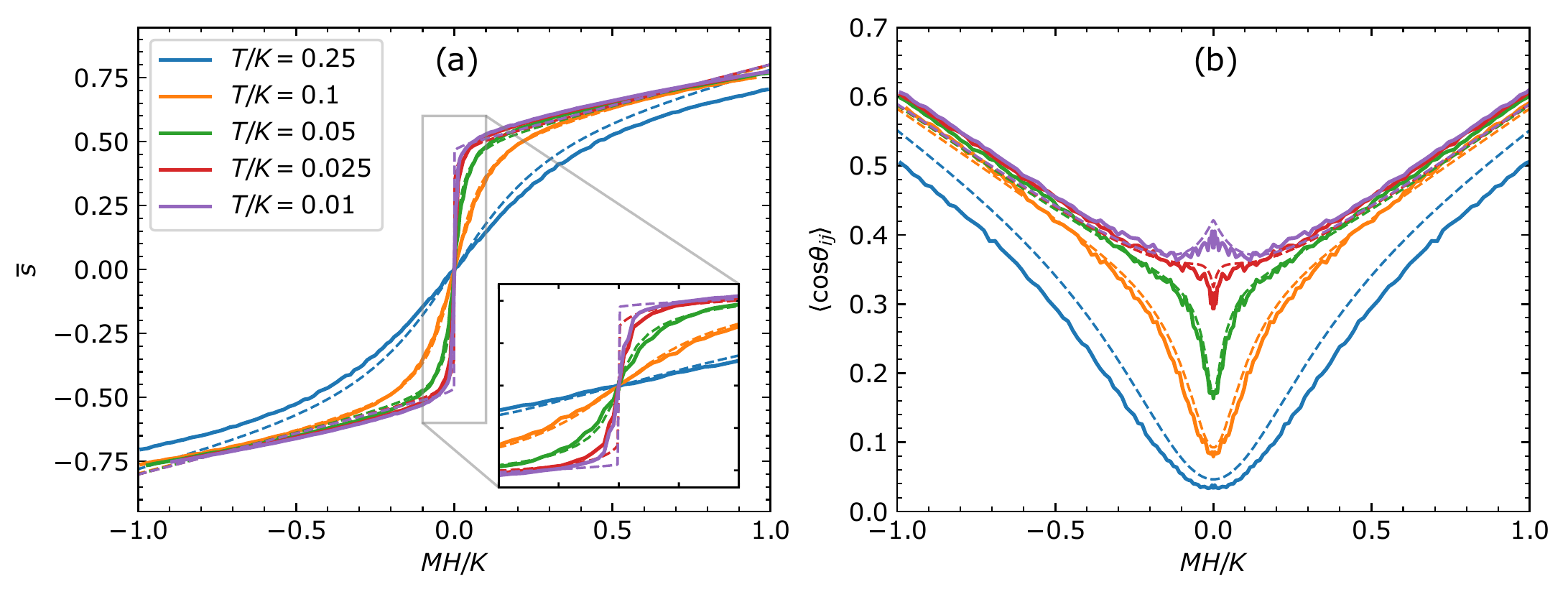}
        \caption{ The comparison between numeric simulation (solid lines) and mean-field theory (dashed lines) for $J = 0.025K$ and different temperatures. (a) the comparison of averaged magnetization, (b) the comparison of $\langle \cos\theta_{ij} \rangle$ that controls TMR.  A phenomenological constant $\alpha = 0.93$ is used in the mean-field theory in both panels. }
    \label{fig:MF-mod}
\end{figure*}

The results of the simulation are presented in Fig.~\ref{fig:MF-mod} for different values of the temperature $T$ in the case of small exchange energy $J=0.025 K$. The magnetization grows very fast at $MH \sim J$. However, its growth slows down when $MH \gtrsim 0.2 K$. One can say that the magnetization reaches quasi-saturation. However, MR has a significant increase up to $MH\sim 1.5 K$. Thus, there is a strong MR with a quasi-saturated magnetization, which will be explained using the mean-field analysis. Also in Fig.~\ref{fig:MF-mod} we compare the results of numeric simulation with the mean-field analysis that is discussed in the next section.

\section{Mean-field analysis}
\label{s:MF}

In this section we describe the mean-field treatment of the model  (\ref{E-gen}). The mean-field theory contains some additional simplification compared to the direct numeric simulation described in Sec.~\ref{s:num}. However, it can be more easily solved numerically and allows analytical solution in some limiting cases. It is compared with numeric simulation in Fig.~\ref{fig:MF-mod}.

\begin{figure}[htbp]
    \centering
       \includegraphics[scale=0.35]{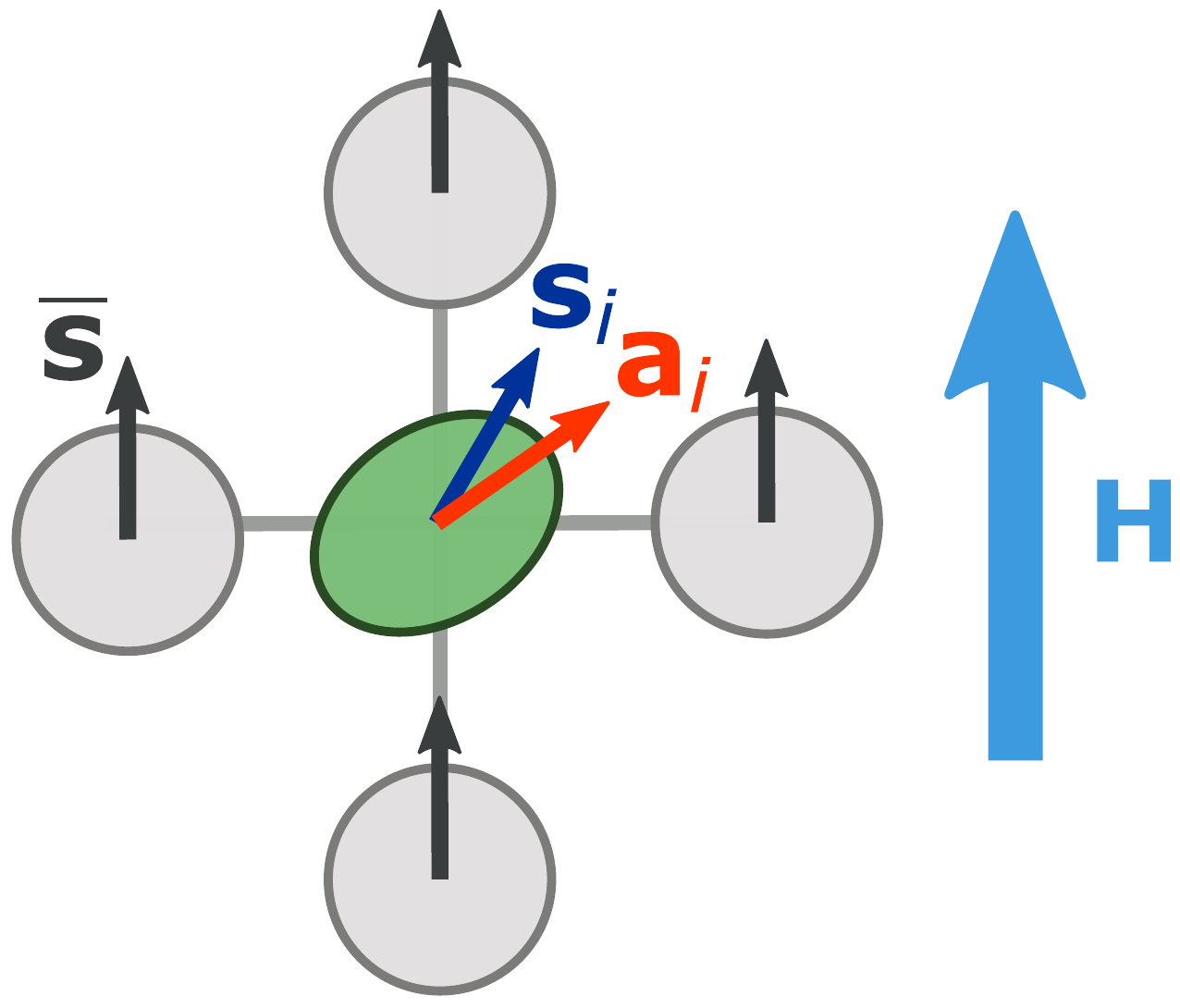}
        \caption{Granule magnetization in the mean-field approximation.  }
    \label{fig:MFpic1}
\end{figure}

Let us consider the granule $i$ connected to four other granules (Fig.~\ref{fig:MFpic1}). The exchange energy related to the granule $i$ is equal to $-J{\bf s}_i \cdot \sum_j {\bf s}_j$. Here index $j$ enumerates granules connected to granule $i$ with the exchange interaction. Within the framework of the mean-field approximation we substitute $\sum_j {\bf s}_j = 4 \overline{\bf s}$ where $\overline{\bf s}$ is the averaged magnetization of the granules. It is directed along the axis of the magnetic field and is equal to $\langle\cos\theta_i\rangle$. It should be found self-consistently.  The exchange interaction  is reduced to the increased effective magnetic field ${\bf H}_{MF} = {\bf H} + 4J\overline{\bf s}/M$. The energy of the granule $i$ in this approximation is $E_i({\bf s}_i) = -K({\bf a}_i{\bf s}_i)^2 - M {\bf s}_i {\bf H}_{MF}$. The average magnetization can be expressed as follows
\begin{equation}\label{si-1}
\overline{\bf s}_i = \frac{\int {\bf s}_i' \exp\left( - \frac{E_i({\bf s}_i')}{T}\right) d{\bf s}_i'}{\int \exp\left( - \frac{E_i({\bf s}_i')}{T}\right) d{\bf s}_i'}.
\end{equation}
Here the integration is taken over all possible unit vectors ${\bf s}_i'$.

When the energy of anisotropy is large compared to the exchange and magnetic energies $K \gg J,MH$, expression (\ref{si-1}) can be significantly simplified because magnetization of the granule is nearly always directed along the easy axis, either in the direction ${\bf a}_i$ or $-{\bf a}_i$. The energy $E_i({\bf s}_i)$ has two minima near these directions, that control the integrals in Eq.~(\ref{si-1})
\begin{multline}\label{si-2}
\overline{s}_i = ({\bf a}_i{\bf e}_H)\tanh \frac{M H_{MF}({\bf a}_i{\bf e}_H)}{T} \\
+ \frac{M H_{MF}}{2K}\left(1 - ({\bf a}_i{\bf e}_H)^2\right).
\end{multline}
Here ${\bf e}_H$ is the unit vector in the direction of the magnetic field. The first term in r.h.s. of Eq.~(\ref{si-2}) describes the thermodynamic distribution of magnetization between minima of $E_i({\bf s}_i)$. The second term shows that the minima are shifted by the effective magnetic field. This term is proportional to $M H_{MF}/K$. The terms proportional to  $(M H_{MF}/K)^3$  are neglected in Eq.~(\ref{si-2}).

The mean-field magnetization can be obtained by averaging of single granule magnetization $\overline{s}_i $  over the easy axis directions of all granules. This averaging yields
\begin{multline}\label{smf-1}
\overline{s} = \frac{1}{2} + \frac{M H_{MF}}{3K} - \frac{\pi^2 T^2}{24 M^2 H_{MF}^2} \\
+ \frac{T \log\left(1 + e^{-2MH_{MF}/T} \right)}{MH_{MF}} \\
- \frac{T^2 {\rm Li}_2\left( e^{-2MH_{MF}/T} \right)}{2M^2 H_{MF}^2}.
\end{multline}
Here ${\rm Li}_s(z)$ is the polylogarithm.
Note that $H_{MF}$ depends on $\overline{s}$. Therefore, Eq.~(\ref{smf-1}) is the equation that should be solved to find $\overline{s}$. It can be done numerically.

In Fig.~\ref{fig:MF-mod}(a) we compare $\overline{s}$ calculated in the mean-field theory with the results obtained by the numeric simulation. The results of the simulation are shown with solid lines and the results of mean-field theory with dashed lines. The colors of the lines correspond to different temperatures. To improve the agreement between mean-field approximation and Monte-Carlo simulation we multiply the magnetization $\overline{s}$, calculated with the mean-field theory, by a phenomenological constant $\alpha$. For small ratios $T/K$ and $J/K$ this constant is close to $1$. The value $\alpha=0.93$ is used in Fig.~\ref{fig:MF-mod}.

The mean-field approximation (\ref{smf-1}) describes the magnetization of the system relatively well. However, in the mean-field theory the magnetization is discontinued at $H=0$. Its behavior in the results of simulation is more smooth.

\begin{figure}[htbp]
    \centering
        \includegraphics[scale=0.35]{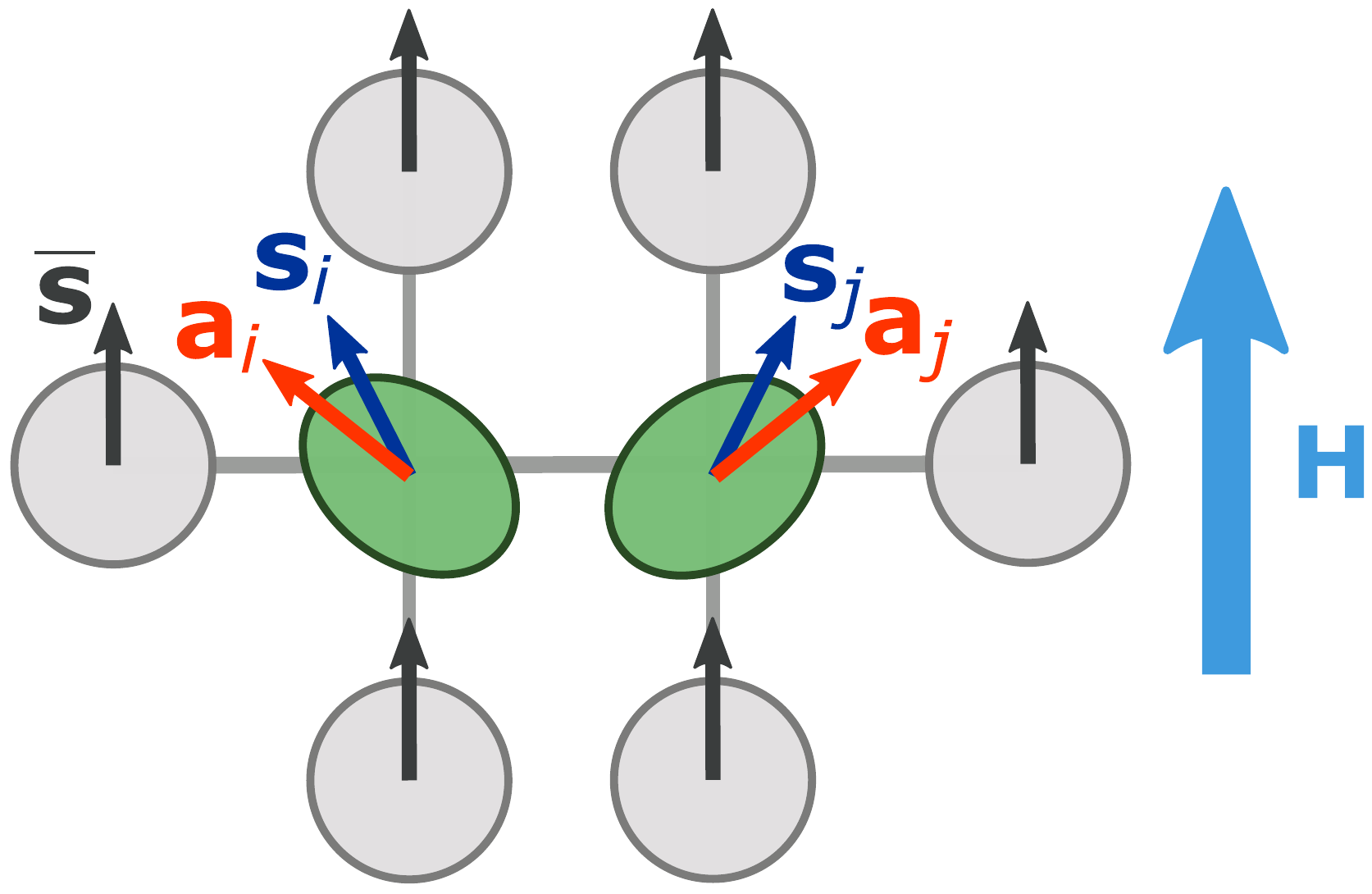}
        \caption{The mean-field scheme for the calculation of averaged $\cos\theta_{ij}$.  }
    \label{fig:MFpic2}
\end{figure}

To describe the relative directions of magnetizations of neighbor granules, it is important to take into account, that even without averaged magnetization $\overline{s}$, the exchange interaction tends to align nearby granules. Therefore,  we modify the mean-field scheme as shown in Fig.~\ref{fig:MFpic2}. We explicitly consider the directions of easy axes and magnetizations of the two granules $i$ and $j$ and treat their 6 neighbors with mean-field approximation.

The energy in this model is equal to
\begin{multline}
E_{ij}({\bf s}_i,{\bf s}_j) = -K ({\bf a}_i{\bf s}_i)^2 -K ({\bf a}_j{\bf s}_j)^2 \\
- M{\bf H}_{MF}'({\bf s}_i + {\bf s}_j ) - J{\bf s}_i{\bf s}_j.
\end{multline}
The mean-field $H_{MF}'$ in this scheme is equal to $H_{MF}' = H + 3J\overline{s}$, because each of the granules $i$ and $j$ is connected with three granules treated with mean-field approximation. The value $\overline{s}$ is taken from the solution of Eq.~(\ref{smf-1}).

In the general case the thermodynamic average of $\cos\theta_{ij}$ is given as follows
\begin{equation}\label{cij-1}
\langle \cos\theta_{ij} \rangle = \frac{\int ({\bf s}_i' {\bf s}_j') \exp\left( - \frac{E_{ij}({\bf s}_i',{\bf s}_j')}{T}\right) d{\bf s}_i' d {\bf s}_j'}{\int \exp\left( - \frac{E_{ij}({\bf s}_i', {\bf s}_j')}{T}\right) d{\bf s}_i' d {\bf s}_j'}.
\end{equation}

With the approximation that is applied in Eq.~(\ref{si-2}) the integration can be reduced to the summation over four minimums of the energy $E_{ij}$. Here we give the expression for $\cos\theta_{ij}$ in the first minimum.
\begin{multline}\label{cs1}
\cos\theta_{ij}^{(1)} = {\bf a}_i {\bf a}_j + \frac{J}{K}(1-({\bf a}_i{\bf a}_j)^2) \\
+ \frac{M{\bf H}_{MF}'}{2K}(1-{\bf a}_i{\bf a}_j)({\bf a}_i + {\bf a}_j).
\end{multline}
The energy of this minimum is equal to
\begin{multline}\label{Es1}
E_{ij}^{(1)} = -({\bf a}_i + {\bf a}_j)M{\bf H}_{MF}' - J ({\bf a}_i {\bf a}_j)\\
- \frac{1}{4K} \Bigl( M{\bf H}_{MF}' + J {\bf a}_j - {\bf a}_i \bigl(
{\bf a}_i \cdot (M{\bf H}_{MF}' + J{\bf a}_j)\bigr) \Bigr)^2 \\
- \frac{1}{4K} \Bigl( M{\bf H}_{MF}' + J {\bf a}_i - {\bf a}_j \bigl(
{\bf a}_j \cdot (M{\bf H}_{MF}' + J{\bf a}_i)\bigr) \Bigr)^2.
\end{multline}
The corresponding expressions for other minima differ from (\ref{cs1}) and (\ref{Es1}) by the inversion of the sign of either one of ${\bf a}_i$ and ${\bf a}_j$ or both their signs.

The expression for the averaged $\cos\theta_{ij}$ that controls the MR is as follows
\begin{equation}\label{cmf1}
\langle \cos\theta_{ij} \rangle = \left\langle \frac{\sum_{n} \cos\theta_{ij}^{(n)} e^{-E_{ij}^{(n)}/T} }{\sum_{n} e^{-E_{ij}^{(n)}/T}}\right\rangle_{a}.
\end{equation}
Here index $n$ enumerate local minima. The averaging $\langle\rangle_a$ is made over all possible directions of the easy axes ${\bf a}_i$ and ${\bf a}_j$.

In the general case expressions (\ref{cs1})--(\ref{cmf1}) can be used to calculate $\langle \cos\theta_{ij} \rangle$ numerically. In Fig.~\ref{fig:MF-mod}(b) we compare $\langle \cos\theta_{ij} \rangle$ calculated with  mean-field theory with numeric simulations described in Sec. \ref{s:num}. Similarly to Fig.~\ref{fig:MF-mod}(a), the results of the mean-field theory are multiplied by the constant $\alpha=0.93$ in Fig.~\ref{fig:MF-mod}(b). With this modification, the mean-field theory is in reasonable agreement with numerical simulations. It means that mean-field approach can give a qualitative understanding of the behavior of the model  (\ref{E-gen}).

In certain limiting cases, it is possible to obtain an analytical expression for the mean-field approximation. When the magnetic energy $MH$ is larger than the exchange energy $J$ and much larger than temperature, it is possible to take into account only the first minimum ($n=1$) in Eq.~(\ref{cmf1}). In this case $\langle \cos\theta_{ij} \rangle$ is described by the averaging of Eq.~(\ref{cs1}) over the directions of easy axes. This averaging leads to the following result
\begin{equation}\label{cmf2}
\langle \cos\theta_{ij} \rangle = \frac{1}{4} + \frac{1}{3}\frac{M H_{MF}'}{K} + \frac{2}{3}\frac{J}{K}.
\end{equation}
Within the same assumptions, Eq.~(\ref{smf-1}) for the averaged magnetization can be reduced to $\overline{s} = 1/2 +MH/3K + 2J/3K$. It leads to
\begin{equation}\label{cmf3}
\langle \cos\theta_{ij} \rangle = \frac{1}{4} + \frac{MH}{3K} + \frac{4}{3}\frac{J}{K}.
\end{equation}
Only the terms linear over $1/K$ are kept in Eq.~(\ref{cmf3}).

\begin{figure}[htbp]
    \centering
        \includegraphics[width=0.45\textwidth]{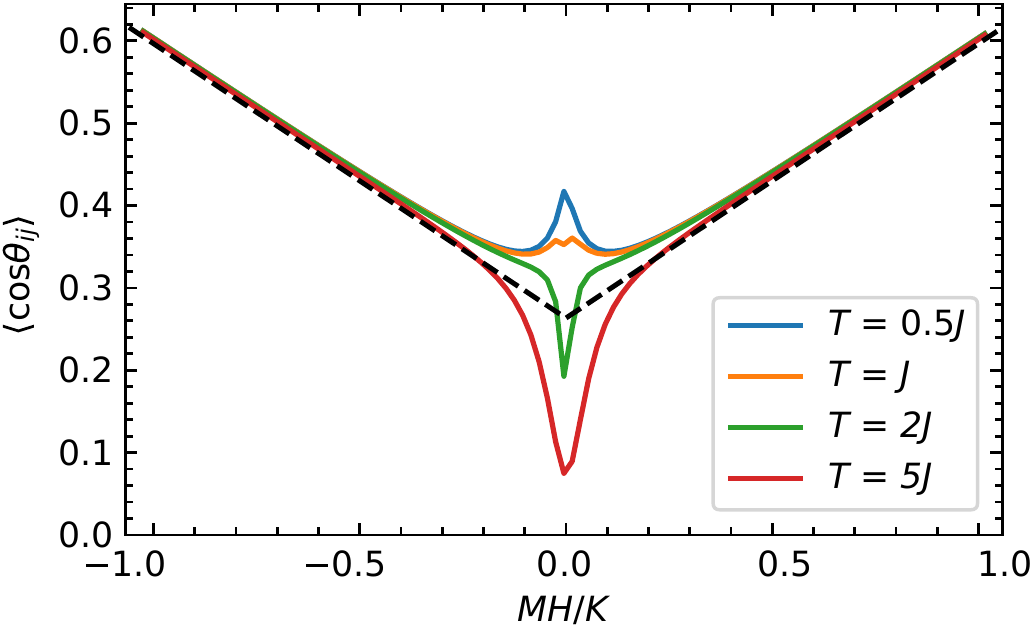}
        \caption{Comparison of Eq.~(\ref{cmf3}) (black dashed line) with the numeric solution of the mean-field model (solid lines) for  $J=0.01K$. Different colors of the solid lines correspond to different temperatures
        from $T=0.5J$ (blue) to $T=5J$  (red). }
    \label{fig:MF-cos2}
\end{figure}

We compare expression (\ref{cmf3}) with the numeric solution of the mean-field model in Fig.~\ref{fig:MF-cos2} for $J=0.01K$. At sufficiently high field Eq.~(\ref{cmf3}) agrees with the numeric solution. It is interesting that at low field the sign of MR can be different and depends on temperature. To analyze these small fields it is useful to neglect $\propto MH/K$ terms and keep only zeroth-order terms over $MH/K$.

When the terms $\propto MH/K$  are neglected in Eqs.~(\ref{cs1}-\ref{cmf1}), $\langle \cos\theta_{ij} \rangle$  is controlled by the relations $MH/J$ and $MH/T$. If we also neglect $J/K$ terms, it allows to reduce Eq.~(\ref{cmf1}) to
\begin{widetext}
\begin{equation}\label{smallK1}
\langle \cos\theta_{ij} \rangle =
\left\langle
({\bf a}_i {\bf a}_j)
\frac{
    \cosh\left(\frac{M{\bf H}_{MF}'({\bf a}_i + {\bf a}_j)}{T}\right) \exp\left(\frac{J {\bf a}_i{\bf a}_j}{T}\right)
    -
    \cosh\left(\frac{M{\bf H}_{MF}'({\bf a}_i - {\bf a}_j)}{T}\right) \exp\left(-\frac{J {\bf a}_i{\bf a}_j}{T}\right)
}{
    \cosh\left(\frac{M{\bf H}_{MF}'({\bf a}_i + {\bf a}_j)}{T}\right) \exp\left(\frac{J {\bf a}_i{\bf a}_j}{T}\right)
    +
    \cosh\left(\frac{M{\bf H}_{MF}'({\bf a}_i - {\bf a}_j)}{T}\right) \exp\left(-\frac{J {\bf a}_i{\bf a}_j}{T}\right)
}
\right\rangle_a.
\end{equation}
\end{widetext}
This equation corresponds to the situation when the magnetization of a granule is always directed along the easy axis. However, one of the two directions along this axis is selected with respect to the temperature, magnetic field and exchange interaction. Eq.~(\ref{smallK1}) contains three energy scales: $MH_{MF}'$, $J$ and $T$. When one of these energies is much larger than the other two, it is possible to find $\langle \cos\theta_{ij} \rangle$ explicitly. At large temperature $T \gg J, MH_{MF}'$ the directions of magnetizations are random and $\langle \cos\theta_{ij} \rangle = 0$. At sufficiently large magnetic field $MH_{MF}' \gg J,T$ magnetization projection on the magnetic field is always positive, in this case $\langle \cos\theta_{ij} \rangle = 1/4$. When the exchange energy is the largest $J \gg MH_{MF}', T$, the magnetization of neighbor granules tries to be directed along each other. Sometimes it leads to the negative projection of the magnetization on the magnetic field. In this case $\langle \cos\theta_{ij} \rangle = 1/2$.

\begin{figure}[htbp]
    \centering
       \includegraphics[width=0.45\textwidth]{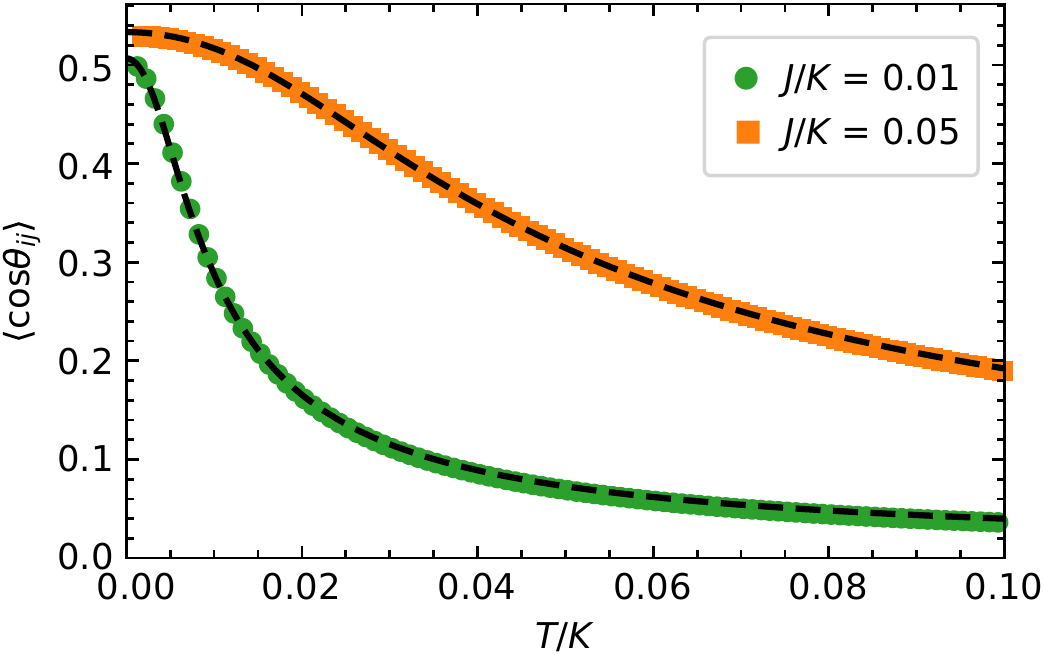}
        \caption{Numeric solution of mean-field model with $H_{MF}' = 0$  (dots) compared with Eq.~(\ref{cmfH0}) (dashed lines). }
    \label{fig:MF-H0}
\end{figure}

To discuss the low-field limit of $\langle \cos\theta_{ij} \rangle$ at the arbitrary relation $J/T$ we consider $H_{MF}' = 0$. Although these value does not appear in the solution of Eq.~(\ref{smf-1}) at sufficiently low temperatures (due to the discontinues magnetization in zero field), it gives qualitatively correct results. In this case Eq.~(\ref{cmf1}) for $\langle \cos\theta_{ij} \rangle$ can be reduced to the explicit expression
\begin{multline}\label{cmfH0}
\langle \cos\theta_{ij} \rangle = \frac{1}{2} + \frac{2J}{3K} - \frac{\pi^2 T^2}{24J^2} +
\frac{T}{J}\log\left(1 + e^{-2J/T} \right) \\
+ \frac{T^2}{2J^2}{\rm Li}_2\left( - e^{-2J/T} \right).
\end{multline}
The term $\propto J/K$ is also included into Eq.~(\ref{cmfH0}). In Fig.~\ref{fig:MF-H0} we compare Eq.~(\ref{cmfH0}) with numeric solution of the mean-field model. A quantitative agreement is achieved.

To understand the sign of the low-field magnetoresistance one should compare $\langle \cos\theta_{ij} \rangle$ described with Eq.~(\ref{cmfH0}) with the value $1/4$ that corresponds to the limit $K \gg MH \gg J,T$. At sufficiently high temperature, zero-field value of $\langle\cos\theta_{ij} \rangle$ is small and tends to zero when $K \gg T \gg J$. External magnetic field increases it to the value $1/4$ and leads to negative magnetoresistance, that is usual for TMR. However, at low temperatures, zero-field value for $\langle\cos\theta_{ij} \rangle$ is larger than $1/4$. It is equal to $1/2$ in the limit $K \gg J \gg T$. External magnetic field decreases it leading to the positive magnetoresistance at low fields. Therefore, in certain cases TMR changes its sign and leads to positive magnetoresistance. Negative magnetoresistance is re-established at higher fields as shown in Fig.~\ref{fig:MF-cos2}.

\section{Comparison with experiment}
\label{s:exp}

 \begin{figure}
 	\centering
 	\includegraphics[width=0.48\textwidth]{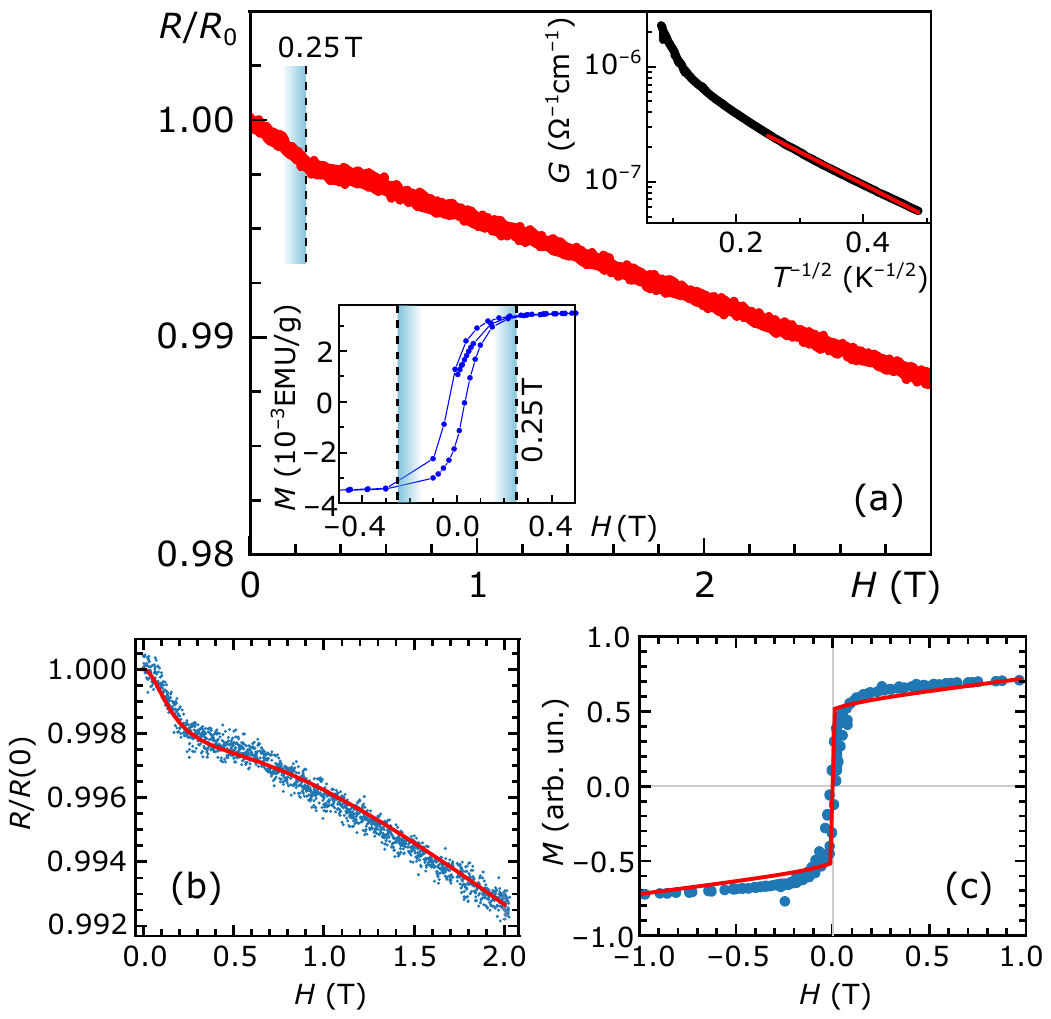}
 	\caption{ (a) magnetoresistance in perpendicular  field  for the SiC${}_x$N${}_y$:Fe film. Top inset: temperature dependence of conductance, bottom inset:  magnetization (data from \cite{ste}).
  (b) comparison of magnetoresistance calculated with the mean-field theory (red line) with experimental data (dots).
 (c) comparison of sample magnetization calculated with the mean-field (red line) theory with the experimental data (dots). }
 	\label{fgr2:mc}
 \end{figure}

In this section we compare the theoretical results with our recent experiments in SiC${}_x$N${}_y$:Fe granular ferromagnetic films.
The SiC${}_x$N${}_y$:Fe films were synthesized using the CVD technique
on high resistance Si(001) substrates by the thermal decomposition
of two different gaseous mixtures.  The detailed growth method, structural characterization, and magnetic properties of obtained  granulated films are described in \cite{ste}. Here we provide the MR data for one typical  sample and compare our experimental results with theoretical data.

In the top inset to Fig.~\ref{fgr2:mc}(a) is the temperature dependence of
conductivity for  the sample  as $G(T^{-1/2})$ plot.
The range of conductivity is too small to reliably determine if $G(T)$ follows Mott or Efros-Shkovskii law. Nevertheless, $G(T^{-1/2})$ dependence in the low-temperature range can be  approximated by the linear law $G(T) \propto \exp[-(T_0/T)^{1/2}]$ with the  value of $T_0$ being about of 2.8 K. It means that it is possible to describe the transport in the sample in terms of VRH.
	
The magnetoresistance curve for this sample is shown in Fig.~\ref{fgr2:mc}(a). The shape of magnetoresistance is quite similar to the one obtained in different granular magnetic films (see the introduction for details). The  curve  includes two regions: low-field ($H \lesssim 0.25{\rm T}$) and high-field ($H \sim 1{\rm T}$) regions. The  range of magnetic fields for low-field region approximately corresponds to the field range of the fast change of magnetization (before it reaches the quasi-saturation, see bottom inset in Fig.~\ref{fgr2:mc}(a)).  In high magnetic fields magnetoresistance  is linear and there is no sign of saturation of the negative magnetoresistance up to the fields $\gtrsim 2 {\rm T}$. This behavior of magnetoresistance is similar to that predicted by our theory. We compare the obtained experimental data with the mean-field model in Fig.~\ref{fgr2:mc}(b) and (c).

In Fig.~\ref{fgr2:mc}(b) we show that the quantitative agreement between our theory (with certain parameters) and experimental data can be reached for magnetoresistance.  Blue points show the measured magnetoresistance while red line corresponds to the mean-field calculation. The parameters taken were as follows: $K = 2.3 M\cdot 1{\rm T}$, $J = 0.19 M \cdot 1{\rm T}$,  $T = 0.13 M \cdot 1{\rm T}$, ${\cal P} = 0.033$.  Here $ M \cdot 1{\rm T}$ is the magnetic energy of the
granule in the magnetic field one Tesla. Therefore, the exchange energy corresponds to the field $0.19
 {\rm T}$ that is similar to the field where magnetization reaches quasi-saturation. Anisotropy energy corresponds to the field $\sim 2.3 {\rm T}$. It means that the negative magnetoresistance can persist up to the fields equal to several Tesla.

In Fig.~\ref{fgr2:mc}(c) we compare magnetization calculated with the mean-field theory with the same parameters and magnetization measured in the SiC${}_x$N${}_y$:Fe film. The mean-field theory cannot capture the hysteresis of magnetization and reduces the low-field behavior of magnetization to the discontinuity at $H=0$. Also, it overestimates the slow increase of magnetization with magnetic field after the quasi-saturation is reached. However, this increase can be observed in the experimental data. Therefore, the quantitative agreement exists between theory and experiment. We believe that qualitative agreement for both the magnetization and magnetoresistance can be reached by taking into account the distributions of granule sizes, anisotropy energies $K$ and exchange energies $J$. However, the discussion of these distributions is out of the scope of this article.

\section{discussion}
\label{s:dis}

We have shown that the exchange interaction between ferromagnetic granules and the magnetic anisotropy with random axis leads to quite sophisticated dependence on TMR on the applied magnetic field. TMR is effectively decoupled from the magnetization and can even change its sign becoming positive at low field. The decoupling of magnetoresistance from magnetization was observed in a number of different granular magnetic materials \cite{Zeise2002,CrO22005,Fe3O42007,Fe3O42018,CoAlN2007, C60Co2010,TiCrN,per2005,CaLaSr,FeMgO,GaAsrev}.  Most often it was ascribed to the so-called ``anti-boundaries''. However, the existence of these ``anti-boundaries'' was not proved independently in most of the discussed materials. We believe that at least in some materials the discussed shape of magnetoresistance can be related not to ``anti-boundaries'' but to the interplay of anisotropy and exchange interaction.

We obtain our results with two methods: with Monte-Carlo numeric simulations and with the mean-field theory. The mean-field theory contains more simplifications that the simulation, however, it was shown that the results of mean-field model are in semi-quantitative agreement with simulation. In some limiting cases the equations of the mean-field model can be solved analytically. In other cases they should be solved numerically, however, their numerical solution is much easier than the Monte-Carlo simulation. Sometimes the mean-field theory allows the analysis of magnetoresistance of the granular systems with parameters that make the converging of the Monte-Carlo calculation problematic.

The shape of magnetoresistance depends on the relations between the anisotropy energy, exchange energy and temperature. Therefore, these parameters of the sample of granular ferromagnetic material can be estimated with the measurements of magnetoresistance. Such an estimate is made for the SiC${}_x$N${}_y$:Fe sample in this work.
However, some care should be taken when comparing the experimental results with the present theory. The model described in this study deals with almost identical granules with the same magnetization, anisotropy energy and exchange interaction between neighbors. Only the easy axes of magnetization are different. However, in real systems there is always some distribution of sizes, exchange energies, etc. For example,  our SEM and HREM data \cite{ste} show the difference in granule sizes and shape and their random distribution in the film.  This distribution can modify the quantitative results of TMR in anisotropic granules with exchange interaction.  Also, to calculate the magnetoresistance, $\langle \cos\theta_{ij} \rangle$ should be averaged not over all the pairs but only over the pairs that are critical for conductivity. Our approach is applicable when there is no correlation between $\cos\theta_{ij}$ and conductivity of the effective resistor connecting granules $i$ and $j$. There is a reason for such a correlation to exist. High conductivity of the resistor $ij$ can be related to large overlap integral between granules $i$ and $j$ that will lead to strong exchange interaction between the granules. Therefore, $\cos\theta_{ij}$ for the granules that are connected with a resistor with high conductivity tends to be larger than the average $\langle \cos\theta_{ij} \rangle$. All these details are not included into the present theory. Nevertheless, even our simplified model shows the possibility to describe decoupling of magnetization
and magnetoresistance in granular arrays without any MR mechanisms beside TMR.

In conclusion, we have shown that TMR in a system of anisotropic ferromagnetic granules with exchange interaction and random easy axes can be decoupled from magnetization. In some cases TMR changes its sign at low magnetic field and becomes positive. We obtained our results with two methods: Monte Carlo simulations and mean-field theory. Our theoretical results agree with our measurements of magnetoresistance in $\rm Fe$ nanocrystals in $\rm SiCN$ matrix.

NPS acknowledge the support from RFBR foundation, grant N 19-42-540001. The work is supported by the Foundation for the Advancement of Theoretical Physics and Mathematics ``Basis''.

\end{document}